\documentclass[10pt,conference]{IEEEtran} 
\IEEEoverridecommandlockouts
\usepackage{amsmath,amssymb,amsfonts}
\usepackage{algorithmic}
\usepackage{graphicx}
\usepackage{textcomp}
\usepackage{xcolor}
\usepackage[a4paper, total={184mm,239mm}]{geometry}
\def\BibTeX{{\rm B\kern-.05em{\sc i\kern-.025em b}\kern-.08em
    T\kern-.1667em\lower.7ex\hbox{E}\kern-.125emX}}

\usepackage{booktabs} 
\usepackage{multirow} 
\usepackage{pgfplots}
\usepackage{subfigure}

\usepackage{balance}

\usepackage{url}
\usepackage{hyperref}

\usepackage{pgfplots}
\pgfplotsset{width=8cm,compat=newest}
\pgfplotsset{every tick label/.append style={font=\small}}

\usepackage[acronym,nonumberlist,nowarn]{glossaries}
\newacronym[longplural={Decision Trees}]{DT}{DT}{Decision Tree}
\newacronym{ASIC}{ASIC}{Application-Specific Integrated Circuit}

\newacronym{HIVE}{HIVE}{HMC Instruction Vector Extensions}
\newacronym{VIMA}{VIMA}{Vector-In-Memory Architecture}
\newacronym{PIM}{PIM}{Processing-In-Memory}
\newacronym{ISA}{ISA}{Instruction Set Architecture}
\newacronym{SSE}{SSE}{Streaming SIMD Extensions}
\newacronym{AVX}{AVX}{Advanced Vector Extensions}
\newacronym{MOB}{MOB}{Memory Order Buffer}
\newacronym{TLB}{TLB}{Translation Look-aside Buffer}
\newacronym{FP}{FP}{Floating-point}
\newacronym[longplural={Functional Units}]{FU}{FU}{Functional Unit}
\newacronym[longplural={Dynamic Random Access Memories}]{DRAM}{DRAM}{Dynamic Random Access Memory}
\newacronym{HPC}{HPC}{High Performance Computing}
\newacronym{HBM}{HBM}{High Bandwidth Memory}
\newacronym{HMC}{HMC}{Hybrid Memory Cube}
\newacronym{CPU}{CPU}{Central Processing Unit}
\newacronym{GPU}{GPU}{Graphic Processing Unit}
\newacronym[longplural={Through-Silicon Vias}]{TSV}{TSV}{Through-Silicon Via}
\newacronym{DDR}{DDR}{Double Data Rate}
\newacronym{NDP}{NDP}{Near-Data Processing}
\newacronym{ML}{ML}{Machine Learning}
\newacronym[longplural={Neural Networks}]{NN}{NN}{Neural Network}
\newacronym[longplural={Convolutional Neural Networks}]{CNN}{CNN}{Convolutional Neural Network}
\newacronym{ReRAM}{ReRAM}{Resistive Random Access Memory}
\newacronym{SIMD}{SIMD}{Single Instruction Multiple Data}
\newacronym{SiNUCA}{SiNUCA}{Simulator of Non-Uniform Cache Architectures}
\newacronym{OrCS}{OrCS}{Ordinary Computer Simulator}
\newacronym{RISC}{RISC}{Reduced Instruction Set Computer}
\newacronym{NFV}{NFV}{Network Function Virtualization}
\newacronym{DU}{DU}{Data Deduplication}
\newacronym{JEDEC}{JEDEC}{Joint Electron Device Engineering Council}
\newacronym{LRU}{LRU}{Least Recently Used}
\newacronym[longplural={Accelerated Processing Units}]{APU}{APU}{Accelerated Processing Unit}

\usepackage[style=ieee]{biblatex}
\begin{document}

\title{Vector In Memory Architecture \\for simple and high efficiency computing\\

}

\author{
\IEEEauthorblockN{
Marco A. Z. Alves\IEEEauthorrefmark{2}
Sairo Santos\IEEEauthorrefmark{2}\IEEEauthorrefmark{3}
Aline S. Cordeiro\IEEEauthorrefmark{2}
Francis B. Moreira\IEEEauthorrefmark{4}
Paulo C. Santos\IEEEauthorrefmark{4}
Luigi Carro\IEEEauthorrefmark{4}
}
\IEEEauthorblockA{
    \IEEEauthorrefmark{2}Department of Informatics -- Federal University of Paraná -- Curitiba, Brazil\\
    \IEEEauthorrefmark{3}Department of Exact Sciences and Information Technology -- Federal Rural University of the Semi-arid -- Angicos, Brazil\\
    \IEEEauthorrefmark{4}Informatics Institute -- Federal University of Rio Grande do Sul -- Porto Alegre, Brazil\\
    Email:\IEEEauthorrefmark{2}\{ascordeiro, mazalves\}@inf.ufpr.br
    \IEEEauthorrefmark{3}\{sairo.santos@ufersa.edu.br\}
    \IEEEauthorrefmark{4}\{pcssjunior, carro\}@inf.ufrgs.br
    }
}

\maketitle


\begin{abstract}
Data movement is one of the main challenges of contemporary system architectures. 
\gls{NDP} mitigates this issue by moving computation closer to the memory, avoiding excessive data movement. 
Our proposal, \gls{VIMA}, executes large vector instructions near 3D-stacked memories using vector functional units, and uses a small data cache to enable short-term data reuse.
It provides an easy programming interface and guarantees precise exceptions.
When executing stream-behaved applications using a single core, VIMA offers a speedup of up to 26$\times$ over a CPU system baseline with vector operations in a single-core processor while spending 93\% less energy.
\end{abstract}

\section{Introduction}

The Von Neumann architecture is the basic design of all modern computers. It requires that any data necessary for computation be placed within the processor before it can be processed. 
This design has been successful for many decades, but as memory technology advancements lagged behind processor technology within the last few years, the "memory wall"~\cite{wulf1995hitting} has become an issue.
Simultaneously, applications that deal with enormous volumes of data have become more common and relevant. 
Since moving data between main memory and CPU incurs high latency and energy, the traditional design falters for such applications \cite{wulf1995hitting, balasubramonian2014near, hashemi2016accelerating}.

Most current computers try to mitigate the latency and energy issues by placing a cache hierarchy close to the processor.
These cache memories store data that has been recently used and may be reused in the future.  
This approach assumes locality and data-reuse patterns, which can benefit many applications that meet such patterns. 
However, this cannot be applied for a set of current data-streaming applications, as they present a non-temporal, data-centric streaming-like  behavior~\cite{xie2018v, qureshi2007adaptive, qureshi2007line, boroumand2018google}. 
Traditional systems present poor speed, and high energy consumption for such applications as the high cost of retrieving data from the main memory becomes unavoidable.
In this scenario, one could argue that prefetchers could solve such a problem. 
Prefetchers would require aggressive policies to exploit full parallelism on the main memory, resulting in massive data movements and cache pollution. 
Thus, such aggressiveness might harm the performance of multiple applications~\cite{eiman2009}.

Aiming to solve this issue, \gls{NDP} is a concept that arises from \gls{PIM}. This idea flips the way the computer deals with data-processing by performing the operations near where the data is stored. 
The recent development of 3D-stacked memories enables \gls{NDP}. 3D-stacked memories are a novel main memory design that stacks up multiple layers of \glspl{DRAM} on top of a layer that features processing capabilities~\cite{HMC2.1}.
Such a design enables the processor to trigger and execute these operations on processing elements inside the memory chip, thus reducing off-chip data movement that would otherwise be inevitable. 
Due to the 3D layout, these memory chips offer high parallelism and low-latency access to the stored data. 
\gls{NDP} proposes placing additional logic processing elements on the logic layer to leverage these desirable capabilities to process the portions of applications that most benefit from them~\cite{lee2016simultaneous}.

A \gls{NDP} design can still follow the von Neumann model by placing complete processors near the data.
This approach, however, may increase complexity and cause temperature issues in a computer system. 
Another possible approach is to extend the model by placing only \glspl{FU} near data, which avoids those issues and allows processors to continue handling tasks they excel at, such as decoding instructions, predicting the outcome of branches, among other functions.

This paper proposes \gls{VIMA}, an \gls{NDP} mechanism that extends the von Neumann model by adding vector functional units near-data, thus avoiding memory-to-processor data movement by performing vector computations near-data. 
Moreover, \gls{VIMA} builds on similar proposed mechanisms~\cite{alves2016large, santos2017operand} by including a small data cache within the memory chip, enabling short-term reuse of vectorized data (i.e., instructions' operands).
\gls{VIMA} advances also include: multi-threading capabilities, easy-to-program interface, extensible design, precise exceptions and interrupts, while being dead-lock free and maintaining high performance and low power consumption.

Our simulation results indicate a speedup of up to 26$\times$ in comparison with traditional x86 architectures for seven evaluated kernels. 
Compared to the state-of-the-art approach, \gls{VIMA} is on average 14\% faster.

\section{Near-Data Processing}

\gls{NDP} has emerged as an accelerator that counterpoints classical architectures.
%
In this approach, a processing element is attached to the same chip as the memory, mitigating data movement between memory and processor.
Thus, \gls{NDP} improves performance and reduces energy consumption by using high parallelism and ensuring low average latency for applications when it highly demands data~\cite{patterson1997case, elliott1999computational}. 
Illustrated on the right side of Figure~\ref{fig.overview}, 3D-stacked memories provide several \gls{DRAM} layers and an additional processing logic layer.
Such scenario exploits the low-latency and high-bandwidth access capabilities provided by the internal parallelism~\cite{mutlu2019enabling, HMC2.1,HMC_Details,Pawlowski_HC} achieved by its multiple independent vaults (similar to channels). 
Compared to \gls{DDR} memory technology, these devices require the same voltage level on average while achieving higher memory bandwidth, reaching up to 320~GB/s~\cite{DDRx, DDR_HBM}, making them more energy-efficient~\cite{HMC_HBM}.

\section{VIMA: Data Reuse Inside the Memory}

\begin{figure}[!hbt]
\centering
\includegraphics[width=1.0\linewidth]{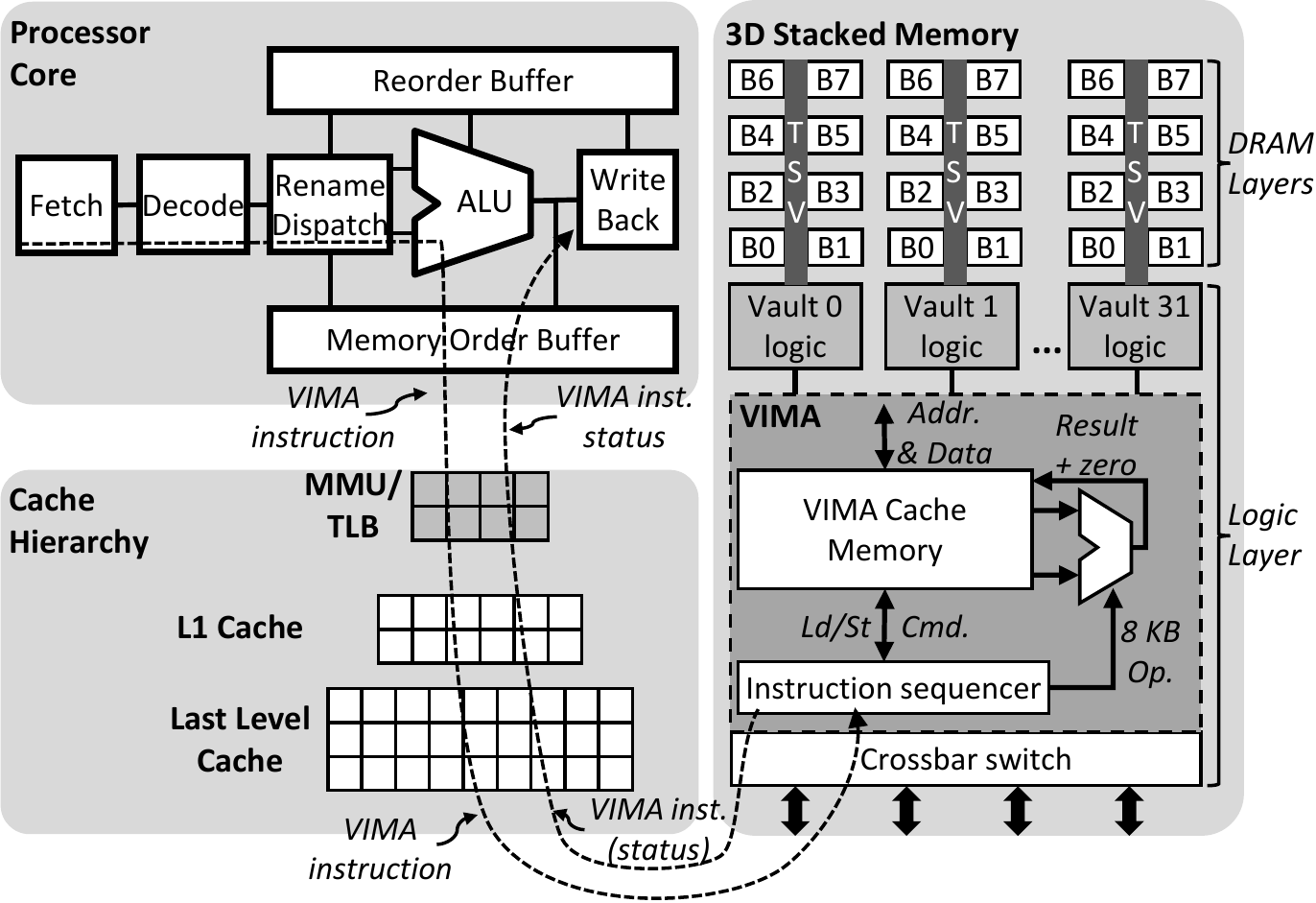}
\caption{3D-stacked memory module with the \gls{VIMA} architecture.}
\label{fig.overview}
\end{figure}

\glsreset{VIMA}
Our proposal is called \gls{VIMA}. 
This architecture adds general-purpose vector operation capabilities to 3D memories to explore the data access parallelism inherent to this architecture.
Figure~\ref{fig.overview} shows \gls{VIMA} inside a 3D-stacked memory. 
The main physical addition of \gls{VIMA} compared to related work is a small cache memory that enables data-reuse of data vectors. At the same time, it maintains most of the performance improvements compared to related PIM strategies. 
One could perceive this cache addition as a minor change. However, \textbf{VIMA enables significant improvements due to its new operation rationale}, as 
    improved short-term data re-usage, 
    easy-to-program interface, 
    precise exceptions, 
    extensible design,
    dead-lock free architecture, and
    multi-threading, 
all discussed in the following sub-sections. 

Similar to other \gls{NDP} approaches, \gls{VIMA} obtains data from several independent memory vaults in parallel~\cite{alves2016large, santos2017operand, tome2018hipe}.
In this paper, we have compared our work with \gls{HIVE} \cite{alves2016large}, given that their full-stack development environment is available.




\subsection{VIMA's Overview}

Like in other vector \gls{ISA} extensions, such as \gls{SSE}, \gls{AVX} and NEON, \gls{VIMA} instructions are inserted into the application by the compiler. 
During execution, \gls{VIMA} instructions traverse the processor pipeline up to the execution stage like a regular memory instruction. 
They are then sent for execution in the 3D-stacked chip, avoiding data movement between memory and processor. 
\gls{VIMA} instructions operate over data vectors of 8~KB.
An 8~KB vector enables the full parallelism of a 3D-stacked memory chip with 32~vaults and at least eight banks per vault.
The small cache memory of 64~KB inside \gls{VIMA} stores up to eight 8~KB vectors enabling \textbf{fast reuse of vector operands}. 
Our flexible design also allows the usage of smaller or larger data vectors, reducing or increasing the parallelism inside the memory, respectively. 
Due to space constraints, this design exploration is out of the scope of this paper.

\subsection{Programming Interface and Binary Generation}\label{intrinsics}
\gls{VIMA} is composed of vector units providing an \gls{ISA} based on ARM NEON vector instructions~\cite{arA57cortex}. Nevertheless, vendors could adapt for their vector \gls{ISA} extensions.
To program for the \gls{VIMA} \gls{ISA} we developed Intrinsics-VIMA, a library inspired by Intel/ARM intrinsics~\cite{lomont2011introduction} available in C/C++ language. 
Any intrinsic library enables low-level code optimization through routine calls written in assembly. 
When the program calls any of these routines, the compiler embeds their \gls{SIMD} instructions directly into the assembly code~\cite{coorporation2009intel, cordeiro2017intrinsics}.
Intrinsics-VIMA routines provide signed and unsigned operations represented in 32 and 64-bit for integer and floating-point single and double precision representations. 

Previous work requires programmers to fine-tune the code to use the available registers or carefully allocate data inside specific memory banks and vaults. 
Intrinsics-VIMA \textbf{simplifies the programming} burden by allowing developers to include \gls{VIMA} function calls on the main code, as it works with any C/C++ library. 
\gls{VIMA} requires existing applications to be adapted and recompiled. Nevertheless, VIMA could easily be supported by  \gls{NDP}-aware compilers such as PRIMO~\cite{ahmed2019compiler}.

\subsection{Processor Required Modifications}\label{modifications}

In our design, the traditional processor triggers the new \gls{VIMA} instructions extension to be executed near-data.
Thus, \gls{VIMA} instructions pass through the pipeline like regular memory instructions. 
All \gls{VIMA} instructions are placed inside the \gls{MOB} and sent into the memory system upon traversing the pipeline. 
Once VIMA returns a signal informing the execution status of an instruction (similar to what happens on \gls{AVX} instructions), the processor reacts to this signal. If it was successfully executed, the processor commits the instruction. Otherwise, it flushes the pipeline, raising an exception and takes the necessary steps to handle it. 

The processor achieves \textbf{precise exceptions} by dispatching one \gls{VIMA} instruction at a time (it dispatches the next instruction after committing the last preceding one). 
This stop-and-go leads to two sources of performance reduction.
First, if \gls{VIMA} uses smaller vector sizes, it will be unable to use the memory's internal parallelism fully. 
For instance, \gls{VIMA} using 256~B vectors performs, on average, 74\% worse than 8~KB.
Besides, the second impact of precise exceptions is the execution gap between instructions. 
The impact of such pipeline bubbles is small for \gls{VIMA}, varying between 2\% and 4\%.
Every \gls{VIMA} instruction generates at least one load or store operation into the memory. Therefore, these memory addresses are translated by the \gls{TLB} and go through permission checks like any memory operation.
We assume hardware support for huge TLB pages, a common feature in modern processors~\cite{kwon2016coordinated}. 
\gls{VIMA} instructions bypass the cache hierarchy. 
The memory addresses touched by any \gls{VIMA} instruction are written back from the system's cache to the main memory before execution to guarantee cache coherence. 
The traditional cache coherence protocol must be \gls{VIMA}-aware.
It must write back dirty lines and invalidate cache belonging to the pages on which \gls{VIMA} will operate.
These instructions then move into the \gls{VIMA} instruction sequencer inside the 3D-memory, where the actual data access happens.


\subsection{VIMA Instruction Sequencing, Vector Units and Cache}\label{sequencing}

\gls{VIMA} requires adding three elements to the 3D-memory: a set of vector functional units, an instruction sequencer, and a cache memory. 
This paper considers a memory with 32 vaults, 8 independent banks per vault, and 256~B row buffer size, but other layouts are also feasible. 
\gls{VIMA} operates over vectors of 8~KB (32$\times$ 256B) utilizing the vault parallelism. One instruction can operate over 2048$\times$ 32-bit elements (e.g., integer) or 1024$\times$ 64-bit elements (e.g., floating-point). 
We used 256 parallel vector units, which means that eight extra cycles are required to fully process the 2048 elements in a pipelined fashion. 
This decision reduces the number of wires between the \gls{VIMA} cache and the vector units. 

\gls{VIMA} executes instructions in-order. 
Instruction executes as soon as the required data is fetched from the memory vaults and available in the \gls{VIMA} cache memory. 
If the data is available in the cache, 1 cycle is required for a tag-check, while another 8 cycles are required for 8 data transfers.
These transfers and the functional units are fully pipelined, enabling complete parallelism.
As most functional units require two operands, our design uses 2 cache ports to complete the operation in 8 cycles (required by the data transfers) plus the number of cycles
required the last pipelined operation.

Once an instruction finishes executing, a status signal is sent to the processor regarding completion or exception (similar to x86 \gls{AVX} instructions).
Before the execution, the sequencer checks the cache for the data that is required by each instruction. 
In case of a cache hit, the operation starts immediately.
The operation ends by writing the results into a fill buffer.
When \gls{VIMA} sends a signal to the processor,
it also writes from this buffer into the cache.
This effectively hides the write operation inside the gap created by the "stop-and-go" approach.
As we write one entire \gls{VIMA} cache line at once, no Read-to-Modify operation is required. 
Whenever the \gls{VIMA} cache evicts this dirty line, it will write the line back to the main memory. 
During a miss, \gls{VIMA} cache uses a \gls{LRU} policy to evict a line. 
VIMA cache splits each vector access into 128 sub-requests to the vault controllers (for 8~KB vectors, considering 64~B cache lines). 
Sub-requests guarantee minimal changes inside the DRAM devices as we are using the same cache line granularity. 
Besides, they are issued to different vaults and banks to increase parallelism. 

Regarding the \textbf{VIMA's cache coherency}, during VIMA's operation, all data are fetched exclusively from this cache. During processor loads, VIMA's cache provides data if available. During processor writes, it writes back and invalidates the respective entry. We consider that such cache could be gated-vdd during long periods of inactivity.


\subsection{Comparison Between VIMA and Related Work}

Most \gls{NDP} related work~\cite{alves2016large, santos2017operand, tome2018hipe}, enables data reuse by using a register bank. 
\gls{VIMA}'s \textbf{design is extensible}, as it can use larger cache memories, while the code does not require modifications to use any extra storage for data-reuse.
Besides, \gls{HIVE} relies on transactions to maintain coherence inside its register bank, which requires locking and unlocking the register bank for each thread during execution, writing \gls{NDP} code as transactions.
Thus, \gls{HIVE} programmers must perform a lock and fetch the necessary data into the register bank before executing operations. 
Moreover, it must also write back all the registers' data to the main memory before unlocking to keep consistency, possibly leading to overheads during loop executions.  
Nevertheless, these additional lock/unlock instructions also overhead the processor pipeline stages and structures.
Meanwhile, \gls{VIMA} does not require any lock/unlock \textbf{eliminating potential dead-lock and process starvation} scenarios, and it performs write-back as needed without a prefixed deadline. 
At the same time, we enable a \textbf{multi-threaded environment} by not locking any structure. 


\begin{figure}[!htb]{
    \begin{tikzpicture}
    \begin{axis}[
        ybar=0pt,
        bar width=6pt,
        xtick=data,
        xtick style= {draw=none},
        height=3cm,
        width=9.5cm,
        ylabel=Speedup,
        y label style={font=\footnotesize,yshift=-2mm},
        x label style={font=\scriptsize},
        y tick label style={font=\scriptsize},
        x tick label style={font=\scriptsize,yshift=-6mm},
        ymax=12,
        ymin=0,
        ytick={0,2,4,6,8,10,12},
        yticklabels={0,2,4,6,8,10,12},
    	ymajorgrids=true,
    	axis line style={draw=none},
    	tick pos=left,
    	enlarge x limits=0.2,
    	enlarge y limits={0.15, upper},
    	nodes near coords,
    	every node near coord/.append style={anchor=west,rotate=90,font=\scriptsize},
    	symbolic x coords={ MemSet, VecSum, Stencil },
    	legend image code/.code={%
    	    \draw[#1, draw=none, fill=black] (-0.01cm,-0.11cm) rectangle (0.31cm,0.11cm);
            \draw[#1, draw=none] (0cm,-0.1cm) rectangle (0.3cm,0.1cm);
        },
        legend style={
            legend columns=-1,
            at={(0.5,1)}, 
            draw={none},
            anchor=south,
            nodes={scale=0.65, transform shape},
            text depth=0pt,
            /tikz/every odd column/.append style={column sep=0cm}},
        legend image post style={scale=0.5},
    	point meta=explicit symbolic
    ]
    
    \addplot [black,/tikz/fill=black!10!white] coordinates {(MemSet,6.06)[6.06] (Stencil,1.97)[1.97] (VecSum,8.15)[8.15]};
    \addplot [black,/tikz/fill=black] coordinates {(MemSet,6.08)[6.08] (Stencil,2.35)[2.35] (VecSum,7.38)[7.38]};
    \addplot [black,/tikz/fill=black] coordinates {\empty};
    \addplot [black,/tikz/fill=black!10!white] coordinates {(MemSet,6.07)[6.07] (Stencil,2.65)[2.65] (VecSum,8.14)[8.14]};
    \addplot [black,/tikz/fill=black] coordinates {(MemSet,6.08)[6.08] (Stencil,2.36)[2.36] (VecSum,7.40)[7.40]};
    \addplot [black,/tikz/fill=black] coordinates {\empty};
    \addplot [black,/tikz/fill=black!10!white] coordinates {(MemSet,6.07)[6.07] (Stencil,2.08)[2.08] (VecSum,8.13)[8.13]};
    \addplot [black,/tikz/fill=black] coordinates {(MemSet,6.08)[6.08] (Stencil,2.5)[2.5] (VecSum,7.41)[7.41]};
    \legend {HIVE, VIMA}
    \end{axis}
    
    \begin{axis}[
        ybar=0pt,
        bar width=6pt,
        xtick=data,
        xtick style= {draw=none},
        height=3cm,
        width=9.5cm,
        y label style={font=\scriptsize,yshift=-2mm},
        y tick label style={font=\scriptsize},
        x tick label style={font=\scriptsize,yshift=-6mm},
        ymax=10,
        ymin=0,
        ytick=\empty,
            yticklabels=\empty,
            xticklabels=\empty,
    	axis line style={draw=none},
    	tick pos=left,
    	enlarge x limits=0.2,
    	enlarge y limits={0.15, upper},
    	nodes near coords,
    	every node near coord/.append style={anchor=east,rotate=90,font=\scriptsize,yshift=3pt},
    	symbolic x coords={MemSet, VecSum, Stencil},
    	legend image code/.code={%
    	    \draw[#1, draw=none, fill=black] (-0.01cm,-0.11cm) rectangle (0.31cm,0.11cm);
            \draw[#1, draw=none] (0cm,-0.1cm) rectangle (0.3cm,0.1cm);
        },
        legend style={
            legend columns=-1,
            at={(0.5,1)}, 
            draw={none},
            anchor=south,
            nodes={scale=0.65, transform shape},
            text depth=0pt,
            /tikz/every odd column/.append style={column sep=0cm}},
        legend image post style={scale=0.5},
    	point meta=explicit symbolic
    ]
    \addplot [black,/tikz/fill=white] coordinates {(MemSet,0) (Stencil,0) (VecSum,0)};
    \addplot [black,/tikz/fill=white] coordinates {(MemSet,0)[8MB] (Stencil,0)[8MB] (VecSum,0)[8MB]};
    \addplot [black,/tikz/fill=white] coordinates {\empty};
    \addplot [black,/tikz/fill=white] coordinates {(MemSet,0) (Stencil,0) (VecSum,0)};
    \addplot [black,/tikz/fill=white] coordinates {(MemSet,0)[16MB] (Stencil,0)[16MB] (VecSum,0)[16MB]};
    \addplot [black,/tikz/fill=white] coordinates {\empty};
    \addplot [black,/tikz/fill=white] coordinates {(MemSet,0) (Stencil,0) (VecSum,0)};
    \addplot [black,/tikz/fill=white] coordinates {(MemSet,0)[32MB] (Stencil,0)[32MB] (VecSum,0)[32MB]};
    
    \end{axis}
    \draw (0,0.1) -- +(0,-0.2);
    \draw (2.53,0.1) -- +(0,-0.2);
    \draw (5.4,0.1) -- +(0,-0.2);
    \draw (7.93,0.1) -- +(0,-0.2);
    \end{tikzpicture}\label{subfig:synth_MatMult_results}
}
\caption{Speedup  of HIVE and VIMA  normalized  to  baseline  AVX  with  a  single  thread (higher is better).}
\label{fig:avx_hive_graphics}
\end{figure}
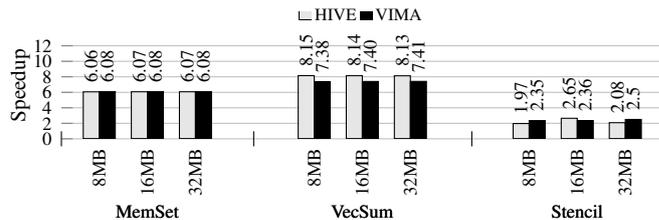

Figure~\ref{fig:avx_hive_graphics} shows the speedup of \gls{HIVE} and \gls{VIMA} compared to a single-threaded x86 baseline, when processing the kernels: \textit{MemSet}, \textit{VecSum}, \textit{Stencil}. 
Results over 1 indicate actual improvement over the baseline. 
Executing \textit{MemSet} on \gls{HIVE} uses the full parallelism inside the 3D-memory by issuing multiple loads in parallel to the register banks. 
However, \gls{HIVE} requires a lock and unlock in the code leading to a sequential write back from the registers to the main memory on every 8 vectors, reducing performance.
\gls{HIVE} executes \textit{VecSum} faster as it better uses the bank parallelism. 
However, such gains come at the cost of non-precise exceptions.
\textit{Stencil} exemplifies how data reuse and the lock-free design can speed up execution, as \gls{VIMA} outperforms \gls{HIVE} in two out of three datasets.

Although VIMA provides vector operations, traditional vector extensions (e.g., AVX, SSE) are still valid for non-data-streaming programs to benefit from the usual cache hierarchy.


\section{Experimental Evaluation of VIMA}

This section presents our methodology and results for \gls{VIMA} evaluation. 
We refer to \gls{AVX} and \gls{VIMA} when discussing applications that use Intel \gls{AVX}-512 and \gls{VIMA}, respectively.
Due to the programming complexity, this section will not bring \gls{HIVE} results.
Please refer to Figure~\ref{fig:avx_hive_graphics} for a direct comparison between \gls{HIVE}, \gls{VIMA} and \gls{AVX} evaluating applications made available by the authors.

\subsection{Methodology}

We used SiNUCA~\cite{alves2015sinuca}, a cycle-accurate simulator, to evaluate the architecture. 
The simulation parameters are similar to Intel's Sandy Bridge microarchitecture. 
Simulation parameter details are disclosed on Table~\ref{tab:config}.

\begin{table}[tb]
    \centering
    \caption{Baseline and VIMA system configuration.}
    \label{tab:config}
    \begin{tabular}{@{}l@{}}
    \hline
    \textbf{OoO Execution Cores} 32 cores @ 2.0~GHz, 32~nm; Power: 6W/core;\\
        6-wide issue; Buffers: 18-entry fetch, 28-entry decode; 168-entry ROB;\\
        MOB entries: 64-read, 36-write; 2-load, 1-store units (1-1 cycle);\\
        3-alu, 1-mul. and 1-div. int. units (1-3-32 cycle);\\
        1-alu, 1-mul. and 1-div. fp. units (3-5-10 cycle);\\
        1~branch per fetch; Branch predictor: Two-level GAs. 4096~entry BTB;\\
    \hline
    \textbf{L1 Data + Inst. Cache} 64~KB, 8-way, 2-cycle; 64~B line; LRU policy;\\
        Dynamic energy: 194pJ per line access; Static power: 30mW;\\
    \hline
    \textbf{L2 Cache} 256~KB, 8-way, 10-cycle; 64~B line; LRU policy;\\
        Dynamic energy: 340pJ per line access; Static power: 130mW;\\
    \hline
    \textbf{LLC Cache} 16~MB, 16-way, 22-cycle; 64~B line; LRU policy;\\
        Dynamic energy: 3.01nJ per line access; Static power: 7W;\\
    \hline
    \textbf{3D Stacked Mem.} 32 vaults, 8 DRAM banks/vault, 256~B row buffer;\\
        4~GB; DRAM@1666 MHz; 4-links@8~GHz; Inst. lat. 1 CPU cycle\\
        8~B burst width at 2.5:1 core-to-bus freq. ratio; Closed-row policy;\\
        DRAM: CAS, RP, RCD, RAS and CWD latency (9-9-9-24-7 cycles);\\
        Avg. energy per access: x86:10.8pJ/bit; VIMA:4.8pJ/bit; Static power 4W;\\
    \hline
    \textbf{VIMA Processing Logic} Operation frequency: 1~GHz; Power: 3.2W;\\
    256 int. units: alu, mul. and div. (8-12-28 cycles for 8~KB pipelined)\\
    256 fp. units: alu, mul. and div. (13-13-28 cycle for 8~KB pipelined);\\
    VIMA cache: 64~KB (8 lines), fully assoc., 2-cycle (1-tag, 1-per data);\\        
    Dynamic energy: 194pJ per line access; Static power: 134mW;\\

    \hline
    \end{tabular}
\end{table}

We used 7 integer and floating-point kernels as workload. 
The integer kernels are \textit{Memory Copy} and \textit{Memory Set}, which generate most of the data movement in big data applications and typical consumer workloads~\cite{boroumand2018google}.
The floating point kernels are \textit{Vector Sum}, \textit{Matrix Multiplication},  \textit{k-Nearest Neighbors}, \textit{Multi-Layer Perceptron}, and \textit{Stencil} convolution, which represent applications like neural networks and computational fluid dynamics processing. 
For all applications (except \textit{MatMul}), we used data sets of 4~MB, 16~MB, and 64~MB. 
We obtained the application traces using Pin~\cite{bach2010analyzing} tool, using VIMA Intrinsics (see section~\ref{intrinsics}).
Below we describe each application:

\noindent \textbf{MemSet}: sets all positions of a vector to a specific value.
\\ \noindent \textbf{MemCopy}: copies the contents of a vector to a new vector in a different memory location.
\\ \noindent \textbf{VecSum}: sums up each element of two input vectors storing the result in an output vector.
\\ \noindent \textbf{Stencil}: convolution using a 5-points stencil over a matrix storing the result in an output matrix.
\\ \noindent \textbf{MatMult}: multiplies two square matrices and stores the results in an output matrix. Due to the long simulation time of this application, we adopted smaller  matrix sizes, resulting in total footprints of 6~MB, 12~MB, and 24~MB.
\\ \noindent \textbf{kNN}: classifies 256 test instances in an n-dimensional space. We used K to equal 9, 32768 training instances, varying the number of characteristics (32, 128, 512).
\\ \noindent \textbf{MLP}: neural network inference step. We used 32768 test instances, varying the number of characteristics (64, 256, 1024). 

\subsection{Results}
In this section we present speedup and energy results of our proposal compared to \gls{AVX} varying the number of threads, and a design space exploration of \gls{VIMA} cache size.

\subsubsection{Single Thread Speedup}

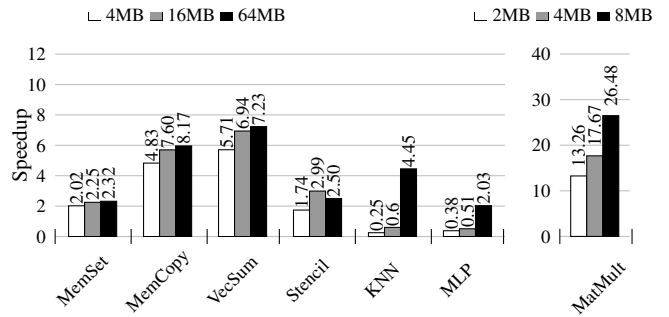
\begin{figure}[!ht]
    \subfigure{
        \begin{tikzpicture}
        \begin{axis}[
            width=7.5cm,
            height=4cm,
            ybar=0pt,
            bar width=6pt,
            xtick=data,
            xtick style={draw=none},
            ytick style={draw=none},
            ylabel=Speedup,
            y label style={font=\footnotesize,yshift=-3mm},
            x label style={font=\footnotesize},
            ymax=12,
            ymin=0,
            axis line style={draw=none},
            ytick={0,2,4,6,8,10,12},
            ytick style={font=\scriptsize},
            yticklabels={0,2,4,6,8,10,12},
            ymajorgrids=true,
            tick pos=left,
            yticklabel style={font=\scriptsize},
            xticklabel style={font=\scriptsize,rotate=45, xshift=-4mm, at={(axis description cs:0.5,-0.1)},anchor=north},
            nodes near coords,
            every node near coord/.append style={font=\scriptsize,rotate=90, anchor=west,xshift=-1mm},
            symbolic x coords={MemSet, MemCopy, VecSum, Stencil, KNN, MLP},
            legend style = { font=\scriptsize, at={(0.3,1.1)}, anchor=south},
            legend image code/.code={
                \draw[#1, draw=none, fill=black] (-0.01cm,-0.11cm) rectangle (0.31cm,0.11cm);
                \draw[#1, draw=none] (0cm,-0.1cm) rectangle (0.3cm,0.1cm);
            },
            legend style = {
                legend columns=-1,
                draw={none},
                text depth=0pt,
                /tikz/every odd column/.append style={column sep=0cm}
                },
            legend image post style={scale=0.5},
            point meta=explicit symbolic
        ]
        \addplot [black,/tikz/fill=white] coordinates {(MemSet,2.02)[2.02]  (MemCopy,4.83)[4.83] (Stencil,1.74)[1.74] (VecSum,5.71)[5.71](KNN, 0.25)[0.25] (MLP, 0.38)[0.38]}; 
        \addplot [black,/tikz/fill=black!40!white] coordinates {(MemSet,2.25)[2.25]  (MemCopy,5.70)[7.60] (Stencil,2.99)[2.99] (VecSum,6.94)[6.94](KNN,0.6)[0.6] (MLP,0.51)[0.51]}; 
        \addplot [black,/tikz/fill=black] coordinates {(MemSet,2.32)[2.32]  (MemCopy,5.96)[8.17] (Stencil,2.50)[2.50] (VecSum,7.23)[7.23] (KNN,4.45)[4.45] (MLP,2.03)[2.03]}; 
        \legend{4MB, 16MB, 64MB};
        \end{axis}
        \draw (0,0.1) -- +(0,-0.2);
        \draw (1,0.1) -- +(0,-0.2);
        \draw (2,0.1) -- +(0,-0.2);
        \draw (2.95,0.1) -- +(0,-0.2);
        \draw (3.95,0.1) -- +(0,-0.2);
        \draw (4.95,0.1) -- +(0,-0.2);
        \draw (5.9,0.1) -- +(0,-0.2);
        \end{tikzpicture}
    \label{subfig:memft_app_results}
    }
    \hspace{-10mm}
    \subfigure{
        \begin{tikzpicture}
    \begin{axis}[
        width=2.5cm,
        height=4cm,
        ybar=0pt,
        bar width=6pt,
        xtick=data,
        xtick style={draw=none},
        ytick style={draw=none},
        ymajorgrids=true,
        ymax=40,
        ymin=0,
        ytick={0,10,20,30,40},
        axis line style={draw=none},
        yticklabels={0,10,20,30,40},
        yticklabel style={font=\scriptsize},
        ytick style={font=\scriptsize},
    	tick pos=left,
    	xticklabel style={font=\scriptsize,rotate=45, xshift=-3.4mm, yshift=-2mm, at={(axis description cs:0.5,-0.1)},anchor=north},
    	x label style={font=\footnotesize},
    	nodes near coords,
    	every node near coord/.append style={font=\scriptsize, rotate=90, anchor=west},
    	symbolic x coords={MatMult, nothing},
        legend style = {font=\scriptsize, at={(0.1,1.1)}, anchor=south},
    	legend image code/.code={%
    	    \draw[#1, draw=none, fill=black] (-0.01cm,-0.11cm) rectangle (0.31cm,0.11cm);
            \draw[#1, draw=none] (0cm,-0.1cm) rectangle (0.3cm,0.1cm);
        },
        legend style={
            text depth=0pt,
            legend columns=-1,
            draw={none},
            /tikz/every odd column/.append style={column sep=0cm}},
        legend image post style={scale=0.5},
    	point meta=explicit symbolic
    ]
    
    \addplot [black,/tikz/fill=white] coordinates {(MatMult,13.26)[13.26]}; 
    \addplot [black,/tikz/fill=black!40!white] coordinates {(MatMult,17.67)[17.67]}; 
    \addplot [black,/tikz/fill=black] coordinates {(MatMult,26.48)[26.48]}; 
    \legend {2MB, 4MB, 8MB};
    \end{axis}
    \draw (0,0.1) -- +(0,-0.2);
    \draw (0.95,0.1) -- +(0,-0.2);
    \end{tikzpicture}
    \label{subfig:memft_matmul_results}
    }
    \caption{Speedup of VIMA normalized to baseline AVX with a single thread (higher is better).}
\label{fig:results_graphics}
\end{figure}

Figure~\ref{subfig:memft_app_results} shows the speedup results for the benchmarks using \gls{VIMA} compared to \gls{AVX} as baseline, while varying the input size. 
Speedup on integer benchmarks \textit{MemSet} and \textit{MemCopy} happens mainly because of the superior use of parallelism in the memory when fetching data without any data reuse. 
Still, it is partially limited because each \gls{VIMA} instruction generates only a single \gls{VIMA} cache miss at a time. 
In contrast, whenever VIMA requires two operands (generating two vector misses), both are requested leveraging the bank parallelism inside each vault.
\gls{VIMA} presents a similar execution time for these two applications, while \gls{AVX} presents faster execution for \textit{MemSet} that have a smaller memory footprint.

The execution of \textit{VecSum} using \gls{VIMA} offers significant performance improvements by making good use of parallelism in the main memory. 
Namely, by fetching two large vectors in parallel, \gls{VIMA} outperforms \gls{AVX} by over 7$\times$ for this benchmark with the largest input size.

The \textit{Stencil} algorithm offers good opportunities for reuse of vectorized data and thus also shows significant speedup. 
Here, data fetches with a single element stride are expected and can be served by the cache. 
Results vary according to input size and matrix size, considering whether the dataset fits inside the last level cache of the baseline system and how efficiently the algorithm deals with different matrix dimensions. 
These factors cause the smaller speedup for the 4~MB input size and the increase of speedup between the 16~MB and 64~MB datasets.

Results for \textit{kNN} and \textit{MLP} using \gls{VIMA} present no speedup whenever the data set used fits in the cache hierarchy of the baseline system (4 and 16~MB cases). 
For these cases, the processor cache hierarchy provides quick access to all the data necessary for processing. 
However, the speedup is considerable when the input size exceeds the size of the last level cache. 
\gls{VIMA} is up to 4x faster than \gls{AVX} for the 64~MB datasets, when the x86 cache presents no help.

The \textit{MatMul} application uses a total of 6~MB, 12~MB, and 24~MB, considering the three matrices. 
For a straightforward, clear comparison of the memory access performance, we used the same algorithm for \gls{AVX} and \gls{VIMA}, which led to higher gains for \gls{VIMA}. 
However, in our tests a tiled algorithm for \gls{AVX} can result in up to 4$\times$ improvements. 
In such scenario, \gls{VIMA} would still over 6.5$\times$ faster for the 24~MB problem size. 



\subsubsection{Multi-Threading Speedup and Energy}

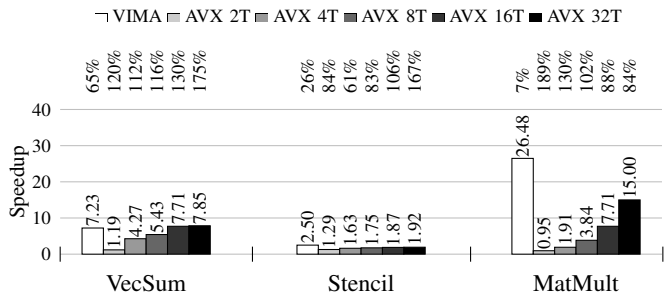
\begin{figure}[!ht]
        \begin{tikzpicture}
        \begin{axis}[
            height=3.5cm,
            width=9.5cm,
            ybar=0pt,
            bar width=8pt,
            xtick=data,
            xtick style= {draw=none},
            ytick style={draw=none},
            ylabel=Speedup,
            enlarge x limits = 0.2,
            y label style={font=\footnotesize,yshift=-2mm},
            x label style={font=\footnotesize},
            ymax=40,
            ymin=0,
            axis line style={draw=none},
            ytick={0,10,20,30,40},
            ytick style={font=\scriptsize},
            yticklabels={0,10,20,30,40},
            ymajorgrids=true,
            tick pos=left,
            yticklabel style={font=\scriptsize},
            nodes near coords,
            every node near coord/.append style={font=\scriptsize,rotate=90, anchor=west,xshift=-1mm},
            symbolic x coords={VecSum,Stencil, MatMult},
            legend style = { font=\scriptsize, at={(0.5,1.5)}, anchor=south},
            legend image code/.code={
                \draw[#1, draw=none, fill=black] (-0.01cm,-0.11cm) rectangle (0.31cm,0.11cm);
                \draw[#1, draw=none] (0cm,-0.1cm) rectangle (0.3cm,0.1cm);
            },
            legend style = {
                legend columns=-1,
                draw={none},
                text depth=0pt,
                /tikz/every odd column/.append style={column sep=0cm}
                },
            legend image post style={scale=0.5},
            point meta=explicit symbolic
        ]
        \addplot [black,/tikz/fill=white] coordinates {(Stencil,2.50)[2.50] (VecSum,7.23)[7.23] (MatMult, 26.48)[26.48]}; 
        
        \addplot [black,/tikz/fill=black!20!white] coordinates {(Stencil,1.29)[1.29] (VecSum,1.19)[1.19] (MatMult, 0.95)[0.95]}; 
        
        \addplot [black,/tikz/fill=black!40!white] coordinates {(Stencil,1.63)[1.63] (VecSum,4.27)[4.27] (MatMult, 1.91)[1.91]}; 
        
        \addplot [black,/tikz/fill=black!60!white] coordinates {(Stencil,1.75)[1.75] (VecSum,5.43)[5.43] (MatMult, 3.84)[3.84]}; 
        
        \addplot [black,/tikz/fill=black!80!white] coordinates {(Stencil,1.87)[1.87] (VecSum,7.71)[7.71] (MatMult, 7.71)[7.71]}; 
        
        \addplot [black,/tikz/fill=black] coordinates {(Stencil,1.92)[1.92] (VecSum,7.85)[7.85] (MatMult, 15.00)[15.00]}; 

        
        
        \legend{VIMA,AVX 2T,AVX 4T,AVX 8T,AVX 16T,AVX 32T};
        \end{axis}
        \draw (0,0.1) -- +(0,-0.2);
        \draw (2.53,0.1) -- +(0,-0.2);
        \draw (5.4,0.1) -- +(0,-0.2);
        \draw (7.93,0.1) -- +(0,-0.2);
        \begin{axis}[
            height=3.5cm,
            width=9.5cm,
            ybar=0pt,
            bar width=8pt,
            enlarge x limits = 0.2,
            axis y line*=right,            
            axis x line=none,
            axis line style = {draw=none},
            xtick=data,
            xtick style= {draw=none},
            ytick style={draw=none},
            enlarge x limits = 0.2,
            ymin=0, 
            ymax=600,
            ytick=\empty,
            yticklabels=\empty,
            ylabel=\empty,
            ymajorgrids=true,
            tick pos=left,
            nodes near coords,
            every node near coord/.append style={font=\scriptsize,rotate=90, anchor=west,xshift=1mm},
            symbolic x coords={Stencil, VecSum, MatMult},
            y label style={font=\scriptsize,yshift=-2mm},
            axis line style={draw=none},
            point meta=explicit symbolic
            ]
            \addplot [only marks] coordinates {(Stencil,600)[65\%] (VecSum,600)[26\%] (MatMult,600)[7\%]};
            \addplot [only marks] coordinates {(Stencil,600)[120\%] (VecSum,600)[84\%] (MatMult,600)[189\%]};
            \addplot [only marks] coordinates {(Stencil,600)[112\%] (VecSum,600)[61\%] (MatMult,600)[130\%]};
            \addplot [only marks] coordinates {(Stencil,600)[116\%] (VecSum,600)[83\%] (MatMult,600)[102\%]};
            \addplot [only marks] coordinates {(Stencil,600)[130\%] (VecSum,600)[106\%] (MatMult,600)[88\%]};
            \addplot [only marks] coordinates {(Stencil,600)[175\%] (VecSum,600)[167\%] (MatMult,600)[84\%]};
        \end{axis}
    \end{tikzpicture}
    \caption{Speedup and energy of VIMA and AVX multithread normalized to baseline AVX with a single thread (higher is better).
    Numbers on the top of the plot present energy consumption relative to AVX single thread.}
\label{fig:results_mt_graphics}
\end{figure}

In comparison to multithreaded \gls{AVX}, our discussion considers only the largest sizes of benchmarks \textit{Stencil}, \textit{VecSum}, and \textit{MatMult}. 
Figure~\ref{fig:results_mt_graphics} compares \gls{VIMA} with an \gls{AVX} implementation using up to 32 cores. 
The percentages above the plot indicate the energy consumption of each execution relative to \gls{AVX} single-threaded execution, in the respective order. 

Considering these results, \gls{VIMA} offers both superior performance and significant energy savings when compared to a single-threaded execution. 
It continues to outperform the baseline system for \textit{VecSum} when these are executed with up to 16 cores in parallel, at a very small fraction of the consumption of energy.
For \textit{Stencil} and \textit{MatMult} applications \gls{VIMA} presents better performance even compared to \gls{AVX} with 32~threads. 
For such applications \gls{VIMA} benefits from the internal 3D-memory parallelism, while the VIMA cache provides necessary data reuse to enable gains. 
At the same time, \gls{VIMA} does not rely on multiple cache levels, which would add extra latency to the memory latencies during a sequence of misses.

\subsubsection{VIMA Cache Size Speedup}

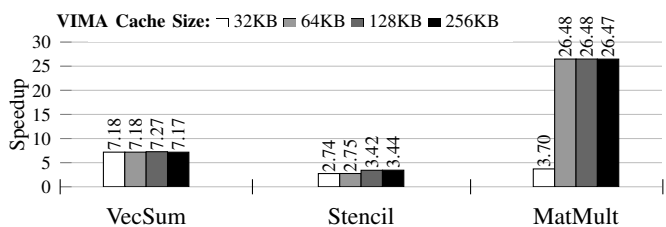
\begin{figure}[!ht]
        \begin{tikzpicture}
        \begin{axis}[
            height=3.5cm,
            width=9.5cm,
            ybar=0pt,
            bar width=8pt,
            xtick=data,
            xtick style= {draw=none},
            ylabel=Speedup,
            enlarge x limits = 0.2,
            y label style={font=\footnotesize,yshift=-2mm},
            x label style={font=\footnotesize},
            ymax=30,
            ymin=0,
            axis line style={draw=none},
            ytick={0,5,10,15,20,25,30},
            ytick style={font=\scriptsize},
            yticklabels={0,5,10,15,20,25,30},
            ymajorgrids=true,
            tick pos=left,
            yticklabel style={font=\scriptsize},
            nodes near coords,
            every node near coord/.append style={font=\scriptsize,rotate=90, anchor=west,xshift=-1mm},
            symbolic x coords={VecSum,Stencil, MatMult},
            legend style = {font=\scriptsize, at={(0.4,1)}, anchor=south},
            legend image code/.code={
                \draw[#1, draw=none, fill=black] (-0.01cm,-0.11cm) rectangle (0.31cm,0.11cm);
                \draw[#1, draw=none] (0cm,-0.1cm) rectangle (0.3cm,0.1cm);
            },
            legend style = {
                legend columns=-1,
                draw={none},
                text depth=0pt,
                /tikz/every odd column/.append style={column sep=0cm}
                },
            legend image post style={scale=0.5},
            point meta=explicit symbolic
        ]
        \addlegendimage{empty legend},
        \addplot [black,/tikz/fill=white] coordinates {(Stencil,2.74)[2.74] (VecSum,7.18)[7.18] (MatMult, 3.70)[3.70]}; 
        \addplot [black,/tikz/fill=black!40!white] coordinates {(Stencil,2.75)[2.75] (VecSum,7.18)[7.18] (MatMult, 26.48)[26.48]}; 
        \addplot [black,/tikz/fill=black!60!white] coordinates {(Stencil,3.42)[3.42] (VecSum,7.27)[7.27] (MatMult, 26.48)[26.48]}; 
        \addplot [black,/tikz/fill=black] coordinates {(Stencil,3.44)[3.44] (VecSum,7.17)[7.17] (MatMult, 26.47)[26.47]}; 

        
        \addlegendentry{\hspace{-.6cm}\textbf{VIMA Cache Size:}};
        \addlegendentry{32KB};
        \addlegendentry{64KB};
        \addlegendentry{128KB};
        \addlegendentry{256KB};
        \end{axis}
        \draw (0,0.1) -- +(0,-0.2);
        \draw (2.53,0.1) -- +(0,-0.2);
        \draw (5.4,0.1) -- +(0,-0.2);
        \draw (7.93,0.1) -- +(0,-0.2);
        
    \end{tikzpicture}
    \caption{Speedup of VIMA with different cache sizes normalized to baseline AVX with a single thread (higher is better).}
\label{fig:results_cache_size_graphics}
\end{figure}

Previous results consider only a fixed VIMA cache size of 64~KB. In this subsection we present how different cache sizes would affect performance of VIMA processing.
Figure~\ref{fig:results_cache_size_graphics} presents speedup results over a single-threaded baseline for the largest dataset of the \textit{Stencil}, \textit{VecSum} and \textit{MatMult} benchmarks. 

We expected small gains for larger VIMA caches as we designed the algorithms to use fewer VIMA cache lines as possible. 
On average, for all seven applications used in this paper, 6 lines would be enough to achieve most of the presented performance. 
However, more complex kernels might benefit from larger VIMA caches, storing temporary data.
Nevertheless, we should be aware that \gls{VIMA} does not intend to replace traditional processors, but provide an efficient accelerator for streaming applications with low data reuse behavior.
\section{Related Work}
Researchers in several distinct areas have explored the \gls{PIM} capabilities of 3D-stacked memories. 
Xie et al.~\cite{xie2017processing} moved a portion of the computations necessary to render 3D images to the logic layer of a 3D-stacked memory. 
The authors aimed to reduce data traffic during some of the more memory-intensive portions of graphics processing algorithms. 
Korikawa et al.~\cite{korikawa2020packet} used \gls{PIM} in \gls{NFV} environments to speed up packet processing by leveraging bank interleaving and channel parallelism of 3D-stacked memories.

Numerous research efforts in \gls{NDP} address big data application requirements, which are particularly susceptible to data movement pitfalls. 
Lee et al.~\cite{lee2019high} identified data redundancy in data centers and proposed a \gls{PIM} accelerator for inline \gls{DU} that significantly reduced latency and power consumption in comparison to traditional \gls{DU} tools. 

Some \gls{NDP} research efforts include placing \glspl{APU} or ARM cores on the logic layer of a 3D-stacked memory. 
These works focus on \gls{ML} training functions~\cite{liu2018processing}, large-scale graph processing~\cite{ahn2015scalable}, and in-memory network frameworks~\cite{gao2015practical}. 
These require ARM cores in conjunction with a \gls{TLB} and rely on routers for vault communication.
\gls{VIMA} is simpler as it does not add cores to the system or rely on vault communication to enhance performance.

Other proposals add specific-purpose cores to the 3D-stacked memory. 
For instance, NIM~\cite{oliveira2017nim}, a reconfigurable \gls{NN} accelerator, like \gls{VIMA}, is a \gls{NDP} architecture that allows for vector operations, features processing units and a sequencer, and is attached to the crossbar switch. 
However, it is much more complex and expensive as it requires one register bank per vault, while \gls{VIMA} stores data in a cache that is accessible from all vaults and enables reuse. 
Several other efforts applied similar principles and offer features like deactivating \glspl{FU} on-demand and enhancing the execution of \glspl{CNN}. However, these required adding modules to each vault of a 3D-stacked memory, making them more expensive and complicated than \gls{VIMA}.

Several \gls{PIM} techniques do not rely on 3D-stacked memories to propose modifying conventional \gls{DRAM} memories and repurposing some of its internal circuits to achieve computation capabilities~\cite{gao2018design, deng2018dracc, li2017drisa, deng2019lacc, sim2018nid, sudarshan2019dram}. 
While this is generally not an expensive approach, it is a lot trickier to program and error-prone than \gls{VIMA} as often programmer must take care of low-level implementation details.
\section{Conclusions}

This paper introduced \gls{VIMA}, a \gls{NDP} mechanism that enables the reuse of data vectors within 3D-stacked memories through a small cache memory.
We have shown that \gls{VIMA} is an efficient approach to reduce data movement, and thus presents itself as an energy-efficient option for a computational system.
\gls{VIMA} improves the execution time of well-known application kernels that present streaming behavior by up to 26$\times$, while also reducing energy consumption by up to 93\%.
Even when executing advanced machine learning application kernels, such as k-Nearest Neighbors and Multi-Layer Perceptron, \gls{VIMA} outperforms traditional processor with AVX512 by up to 6$\times$.

In our analysis with stream-like applications, traditional processors require, on average, 16~cores to reach the same performance level of \gls{VIMA}. 
Thus, \gls{VIMA} efficiently provides high performance consuming a fraction of \gls{AVX} energy.


In the future, we would like to expand the usage of \gls{VIMA} on applications with non-spatial data locality. Planning also a compiler pass for automatic conversion of AVX into \gls{VIMA} instructions, creating a transparent programming interface. 





\balance 
\printbibliography

\end{document}